# REDUCING CERTIFICATION GRANULARITY TO INCREASE ADAPTABILITY OF AVIONICS SOFTWARE

*Martin Rayrole, David Faura, Marc Gatti, Thales Avionics, Meudon la Forêt, France*

## Abstract

A strong certification process is required to insure the safety of airplanes, and more specifically the robustness of avionics applications.

To implement this process, the development of avionics software must follow long and costly procedures. Most of these procedures have to be re-executed each time the software is modified.

In this paper, we propose a framework to reduce the cost and time impact of a software modification. With this new approach, the piece of software likely to change is isolated from the rest of the application, so it can be certified independently. This helps the system integrator to adapt an avionics application to the specificities of the target airplane, without the need for a new certification of the application.

## Introduction

The Integrated Modular Avionics (IMA) principles offers the possibility to host several avionics software applications in a single avionics module, by using partitioning properties specified in avionics standards such as ARINC 653 [1]. These principles reduce the place, the weight and the consumption of avionics equipments, but they increase the interference between applications.

This interference makes the certification more complex, because one has to prove that an application failure cannot be propagated to another one on the same module. This additional certification work is significant because, as presented in [2], the impact of IMA in the certification process concerns most of the functional domains of an avionics platform.

In this paper, we propose a framework to reduce this additional certification work. The proposed framework allows isolating a piece of software from the rest of the application. At run time, a mechanism guaranties a strong (space and time) partitioning between these two parts of the application. These properties offer the possibility to modify and re-certify the isolated piece of software, without re-certifying the rest of the application.

## A New Approach for Building Avionics Applications

With the "state of the art" technologies, the certification of software obliges to freeze the entire source code before delivering the software to a system integrator. No additional source code can be added to certified software without re-launching the certification process for this software. To deal with this constraint, two solutions are used to adapt a software behavior to the integration context: defining configuration data, and using incremental certification.

### *Configuration Data*

Some configuration data can be defined by the avionics application supplier to select the software behavior among a set of pre-defined behaviors.

Configuration data are easy to use by the system integrator, but they are limited to tuning capabilities that have been foreseen in detail during the software development phase.

### *Incremental Certification*

The incremental certification is an efficient solution to deal with this certification complexity of IMA. The incremental certification mechanisms allow to independently certify the avionics platform and each of the partitions hosted in this platform. Thanks to these mechanisms, one partition can be modified in order to be adapted to a particular context, without affecting the certification of other partition housed in the same calculator.

The properties of incremental certification can be used to isolate two sub-parts of an avionics application: the piece of software likely to change can be run in an additional partition. The application is then composed of two partitions that can communicate through inter-partition communication mechanisms.

The creation of an additional partition offers good capabilities to adapt the software behavior, but it is heavy to implement and not adapted to small pieces of code.

### *Proposed Approach*

The proposed approach offers wide capabilities to adapt an avionics application, with a minimum impact on the certification work.

For that, a new framework is defined. This framework offers the capability to modify a software behavior by adding a piece of source code, which can be certified independently from the software.

This source code is compiled with a certified macro-compiler, and dataloaded on the embedded avionics calculator (called the "target platform"). During the target platform startup procedure, the compiled file is checked to verify its compatibility with the platform configuration. At run time, the compiled file is read, interpreted and executed inside a container, which guaranties the strong memory and time segregation between the compiled file execution and the other platform activities.

In the next chapters, we will present the proposed framework. Then we will describe the two main tools that support the framework: a "macro-compiler" and a "container". Finally, we will show how the framework can be used in a use case example.

## Framework Description

### *Framework actors*

Four actors are involved in the proposed framework:

- The "Module Supplier" provides the target platform. It also provides the couple of tools (a "macro-compiler" and a "container") that supports the framework.
- The "Function Supplier" provides the avionics application. This application must be designed to be integrated within the framework.
- The "Adaptation Supplier" provides the piece of code that implements an adaptation of the application. This is a new actor with regard to the existing avionics integration processes.
- The "System Integrator" provides the platform configuration. This configuration defines the resources allocated to the avionics application and to the container.

The figure 1 gives an overview of the proposed framework, with the components provided by each actor.

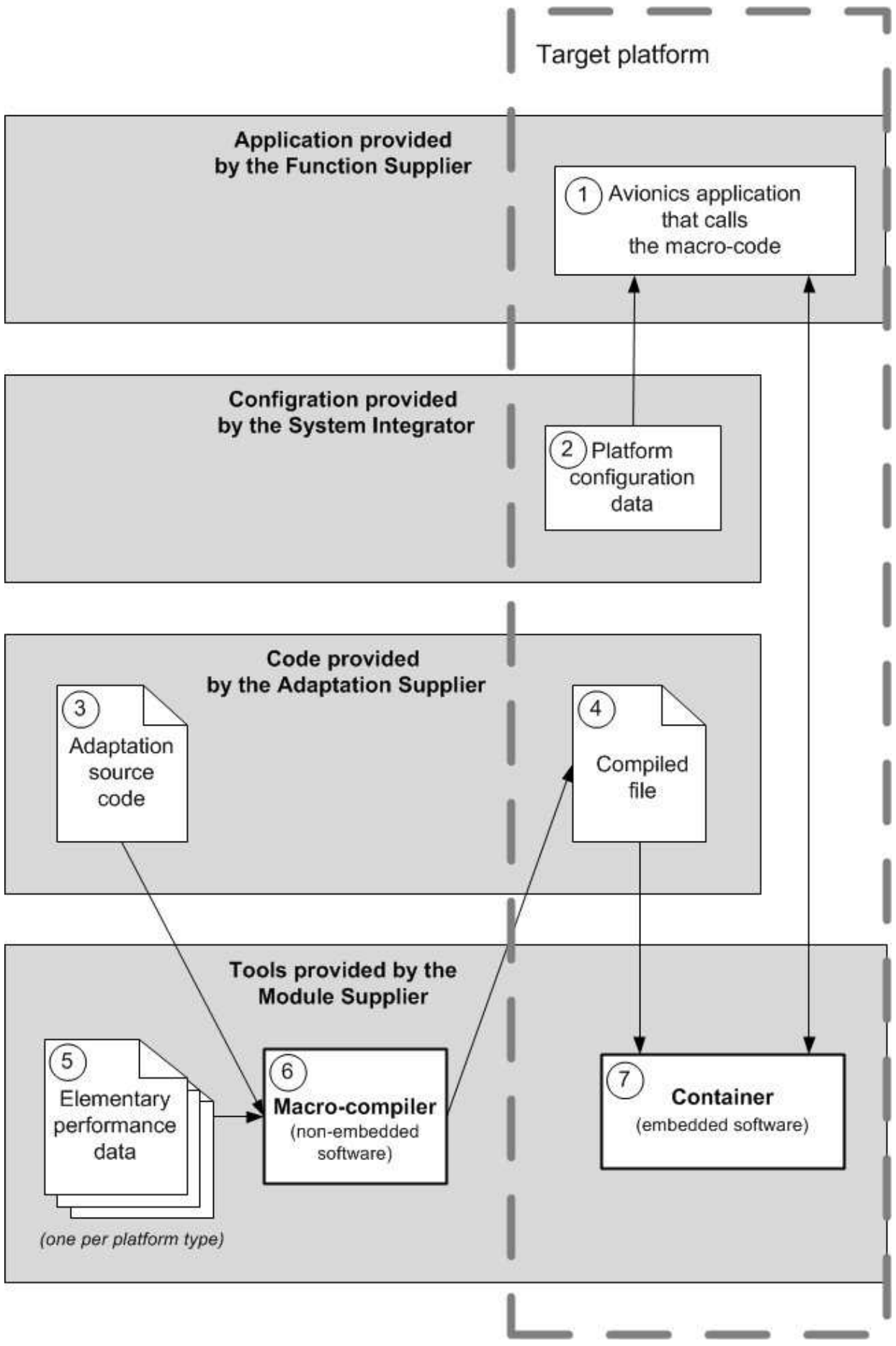

**Figure 1. Framework Overview and Involved Actors**

Depending on the context, the "Adaptation Supplier" activities are supported by one of the three other actors.

The Adaptation Supplier activities are typically supported by the System Integrator, to adapt the application behavior to the specificities of the airplane architecture. However, the Adaptation Supplier activities may also be supported by the Function Supplier to adapt a generic application to its clients' needs. It may also be supported by the Module Supplier to adapt a generic platform to its clients' needs or to hardware specificities.

### *Framework Usage Scenario*

In this paragraph, the numbers in parentheses refers to the numbers in figure 1.

The Function Supplier first provides an application (1) that realizes an avionics function. This application has been designed to be integrated in the framework, which means that a part of its functionalities is supposed to be developed by an Adaptation Supplier.

The Adaptation Supplier writes its own piece of source code (3) that implements an adaptation of the application behavior. He then compiles his source code by using the macro-compiler (6) provided by the Module Supplier. A set of "Elementary performance data" files (5) is also provided by the Module Supplier and used as an input to the macro-compiler.

The output of the macro-compilation is a "compiled file" (4) that is sent to the System Integrator.

The System Integrator uploads the compiled file on the target platform, as well as the application code, the platform configuration data (2), and the container (7). The container is a piece of software provided by the Module Supplier that can execute a compiled file without interfering with other platform activities.

At runtime, the application reads configuration data to know the amount of resources allocated to it. Then, each time it is needed, the application requests the container to execute the compiled file.

### *Supporting Tools Certification*

The macro-compiler and the container must be developed according to DO-178 [3] standard, and certified at least to the same Design Assurance Level (DAL) as the target platform.

## Macro-Compiler Functionalities

The macro-compiler generates a compiled file from an adaptation source code and elementary performance data files, as shown in figure 2.

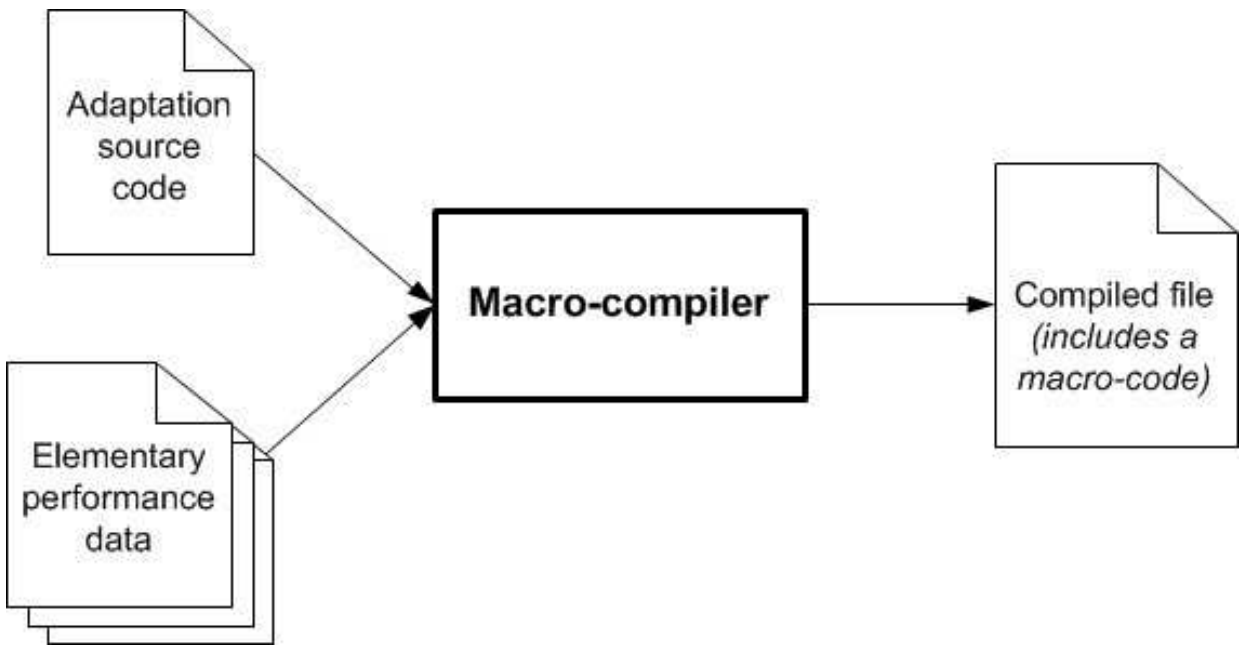

**Figure 2. Macro-compiler Inputs and Output**

The compiled file contains, among other information, a "macro-code" that is the result of the compilation of the adaptation source code. The instructions composing the macro-code are taken from a set of instructions independent from the target platform.

### *Adaptation Source Code (Macro-Compiler Input)*

The adaptation source code respects a programming language syntax defined by the Module Supplier. This programming language is independent from the target platform and from the application.

This programming language has some specific properties that are detailed in the following paragraphs.

#### WCET calculation

The programming language must allow an automatic calculation of the WCET of the macro-code. Several techniques exist for calculating a WCET. A classification of these techniques has been proposed in [4].

The capability to automatically calculate the WCET may have some impacts on the programming language syntax. For example, the maximum number of loop iterations may be explicitly specified in the source code. The macro-compiler could also implement a method, like in [5], to bound the maximum number of loop iterations without an explicit declaration.

**Variables Visibility**

The programming language must distinguish between external and local variables.

External variables are visible inside the adaptation code, and inside the calling avionics application. External variables are typically used as a communication mean between the adaptation code and the calling application. They can be used by the calling application to give input data to the adaptation code, and by the adaptation code to return data to the calling application.

Local variables are visible only inside the adaptation code.

**Memory Access Control**

The programming language must allow a strict control of the memory access. This control guaranties that the adaptation code will never access memory outside the defined variables.

A simple solution to offer this guaranty is to forbid the usage of pointers, and to add the detection of out of bound array access.

### *Elementary Performance Data (Macro-Compiler Input)*

The macro-compiler needs some performance data to calculate the WCET of the macro-code.

These data are stored in "Elementary Performance Data" files that are given as input to the macro-compiler.

Each file contains performance data corresponding to one platform type. A "platform type" represents all target platforms on which the execution of a compiled file takes exactly the same time. In the proposed framework, a "platform type" is characterized by

- The type and the version of the platform hardware.
- The type and the version of the Operating System.
- The version of the container.

These files contain, for each platform type, the WCET of each instruction of the macro-code instruction set. Each WCET value represents the worst case time for reading, interpreting and executing the instruction on the platform type.

### *Compiled File (Macro-Compiler Output)*

The compiled file generated by the macro-compiler is usable for a pre-defined set of platform types. This set of platform types is defined by the set of elementary performance data files given as input to the macro-compiler.

The generated compiled file contains

- A header that gives some characteristics of the compiled file, including the memory size needed to execute the macro-code. The exact content of this header is described below.
- The macro-code, which is the translation of the adaptation source code into a list of instructions. These instructions are taken from a set of elementary instructions independent from the platform type.
- The list of platform types able to execute the macro-code, and the macro-code WCET for each platform type. This WCET represents the worst case time for interpreting and executing the macro-code inside the container. This WCET also includes the time needed to analyze the request sent by the calling application.

**Compiled File Header**

The compiled file header contains all the information required by the container to parse the file, and all the information needed to guaranty the strong time and memory partitioning of the macro-code execution.

For that, the header contains at least the following information:

- The type and version of the macro-compiler, which determines the exact format of the compiled file.
- The macro-code length.
- The memory size needed for external variables (with no padding bytes).
- The memory size needed for local variables (with no padding bytes). The distinction between external and local variables is important because, as explained later (see "**Erreur ! Source du renvoi introuvable.**" paragraph below),

- external and local variables are stored in two different memory spaces.
- The number of platform types considered by the macro-compiler for this adaptation code. This number matches the number of platform type and WCET stored at the end of the compiled file.

## Container Functionalities

The container is in charge of reading the compiled file, and of interpreting and executing the macro-code, when requested by the calling application. These operations are done with two guaranties:

- The execution time of the macro-code execution cannot exceed the WCET stored in the compiled file.
- The accesses done by the container are limited to the container's data and external variables.

### *Container Memory Mapping*

At runtime, the container is an embedded piece of software, running in the same partition as the avionics application that will request the execution of the macro-code.

The container needs to access the following memory spaces:

- The macro-code to execute,
- The local variables defined in the macro-code,
- The external variables defined in the macro-code,
- The container's code that receives and executes requests from the calling application,
- Some additional data internally used by the container's code

All these memory spaces are contained inside the memory allocated to the container, except the external variables. These external variables are used as a communication mean between the calling application and the container and, for that purpose, are shared between these two pieces of software.

In the proposed framework, the external variables are stored in the memory allocated to the calling application. So the container must have access to this memory, but the container behavior guaranties that it will never access to the calling application's data outside of these external variables. This guaranty is provided by the programming language properties and the macro-compiler (see "Memory Access Control" paragraph above) and by the container itself during the application start-up procedure (see "Initialization Phase" paragraph below).

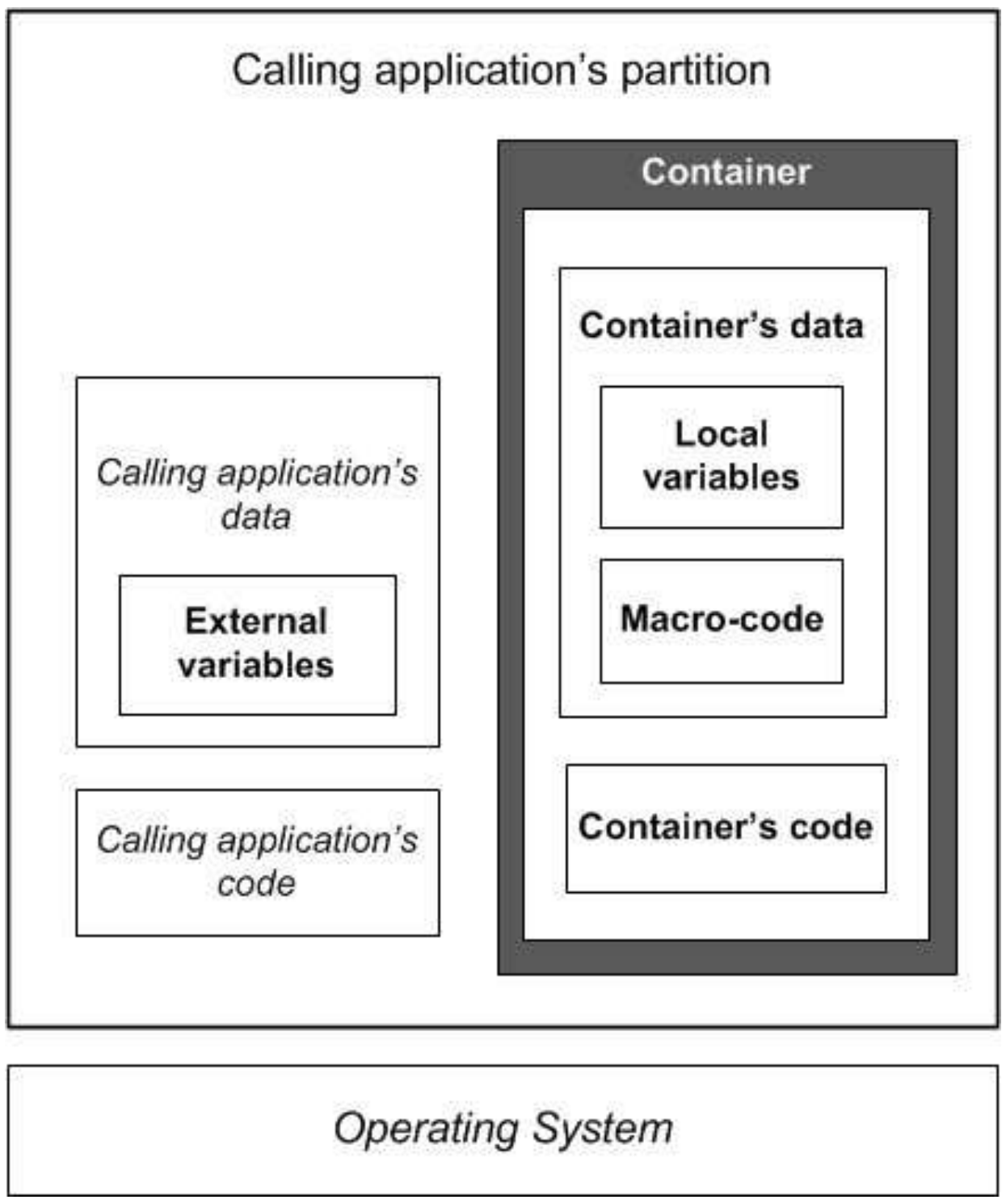

**Figure 3. Container Memory Mapping**

### *Sequence Diagrams*

The calling application sends a request to the container during the application start-up phase, to initialize the macro-code execution. Then, the calling application sends a request to the container, each time it is needed, to execute the macro-code.

These two phases are presented in the next paragraphs.

#### Initialization Phase

The figure 4 shows the sequence diagram of the initialization phase.

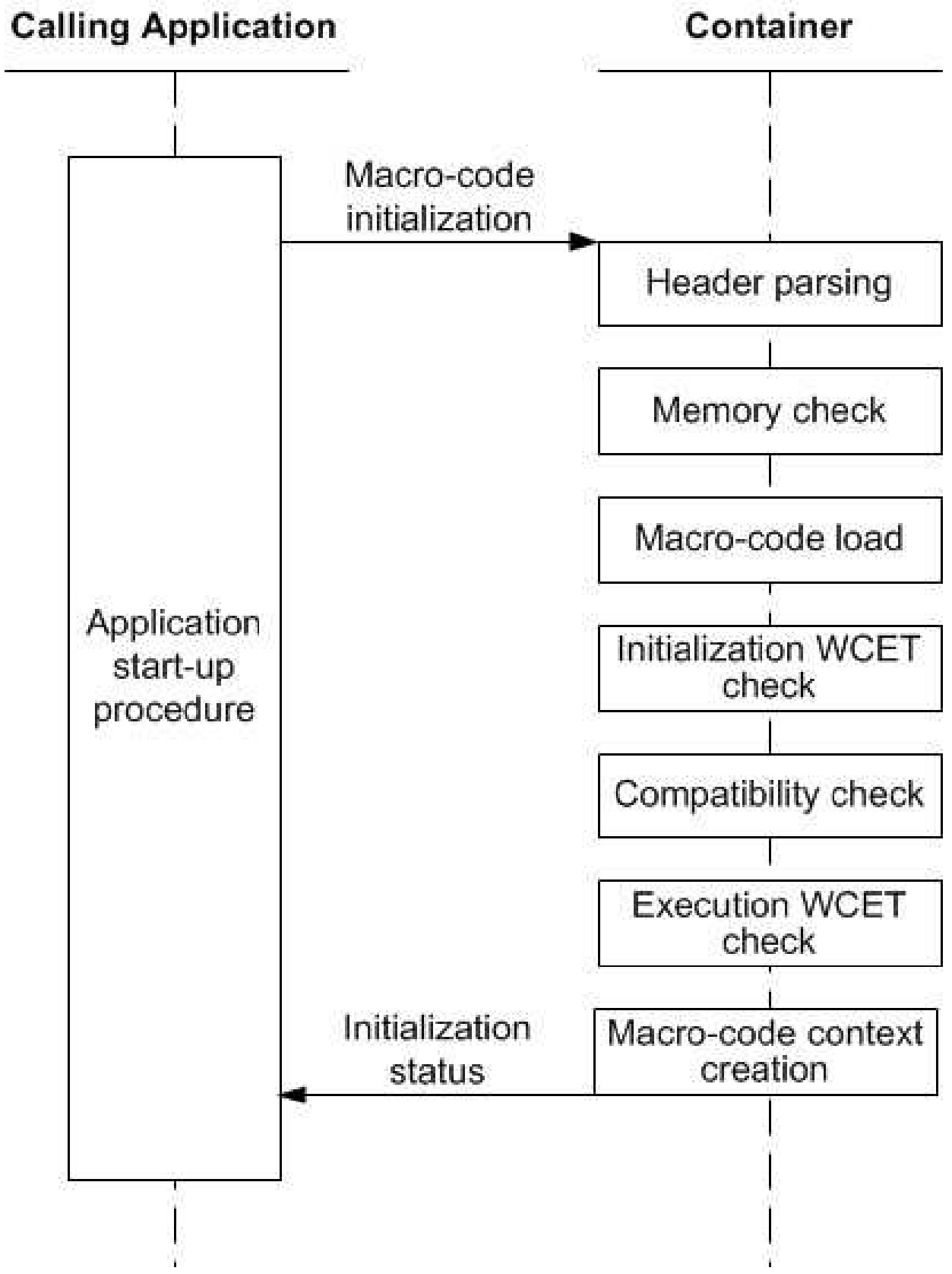

**Figure 4. Initialization Phase Sequence Diagram**

During its start-up procedure, the calling application requests a macro-code initialization to the container. The main objective of this initialization is to detect any configuration error that could prevent the correct execution of the macro-code. The initialization phase is also used to make the macro-code ready to be executed.

From a functional point of view, the calling application could request the initialization of the macro-code just before its first execution. However, for safety considerations, it is most recommended to initialize the macro-code during the application start-up procedure. This recommendation avoids detecting a macro-code incompatibility during operational mode, where failures are much more critical.

The macro-code initialization request contains the following parameters:

- The compiled file address in non-volatile memory.
- The starting and ending memory addresses of the external variables.
- The allocated time for the macro-code execution.

All these parameters are calculated by the calling application from configuration data, which include the amount of resources allocated to the application's partition.

When receiving this request, the container executes the following tasks:

- Header parsing.

  The header of the compiled file is read from the non-volatile memory, and its content is parsed.
- Memory check.

  The memory sizes for macro-code and local variables (in the compiled file header) are compatible with the memory allocated to the container. And the memory size for external variables (also in the compiled file header) is exactly equal to the memory size received in the request.
- Macro-code load.

  The macro-code is copied from the non-volatile memory to the Random Access Memory (RAM).
- Initialization WCET check.

  The number of platform types contained in the compiled file header is checked to be less or equal to the maximum number of platform types configured for the container. This verification is needed to bound the execution time of the initialization phase.
- Compatibility check.

  The platform on which the container is currently running appears in the list of platform types stored in the compiled file.
- Execution WCET check.

  For the current platform type, the WCET stored in the compiled file is less or equal to the allocated time received in the macro-code initialization request.
- Macro-code context creation.

  A macro-code context is created to memorize all data needed to launch the macro-code execution: the memory addresses for the macro-code, for the external variables, and for the local variables.

Finally, the container sends a response to the calling application with the initialization status and, if it is successful, with a macro-code context identifier.

**Execution Phase**

The figure 5 shows the sequence diagram of the execution phase.

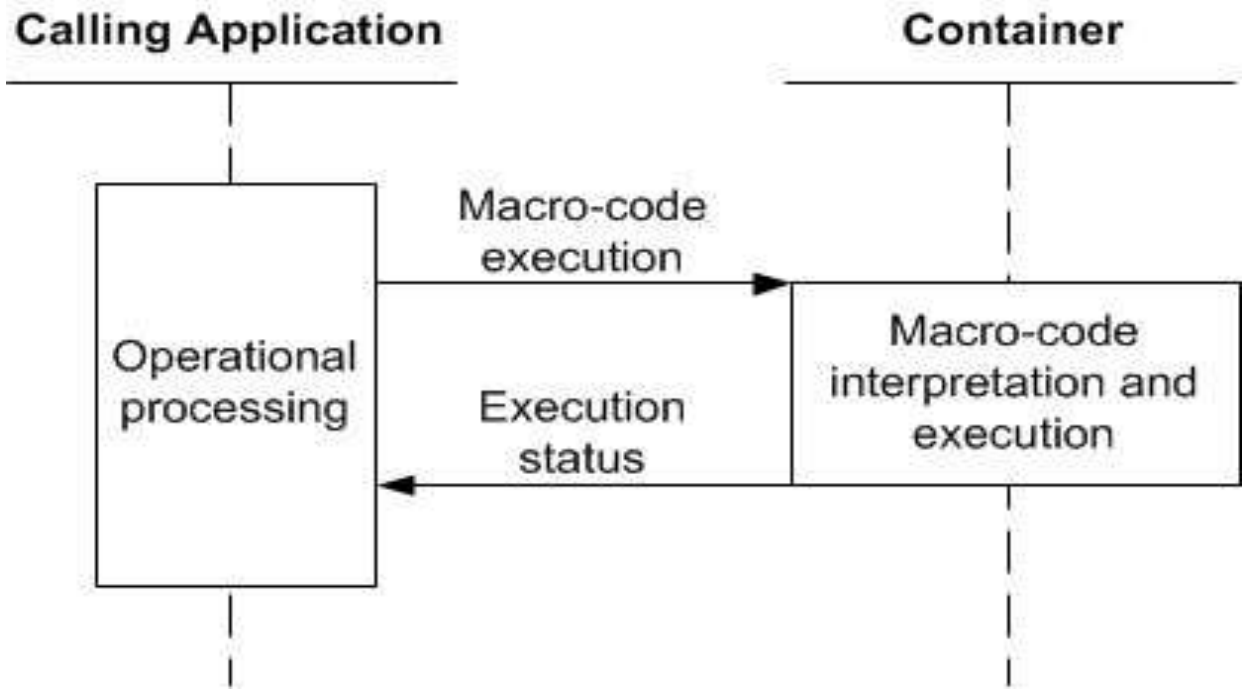

**Figure 5. Execution Phase Sequence Diagram**

Once the macro-code has been initialized, the calling application can request its execution. Only the macro-code context identifier needs to be sent in the execution request.

When receiving an execution request, the container interprets and executes the macro-code. At the end of the execution, a response message is returned to the calling application, with the macro-code execution status. The calling application can then access the external variables to read the result values of the macro-code execution.

## Multi Macro-Code Support

The presented framework is compatible with the use of several macro-codes in one calling application.

In such a case, a compiled file is generated for each adaptation source code, and all these compiled files are dataloaded on the target platform.

During its start-up procedure, the calling application requests the initialization of each macro-code. The container creates a macro-code context for each.

Once in operational state, the calling application can request the execution of any of the macro-codes, independently one from each other.

It should be noticed that the Adaptation Supplier activities, for each source code, can be supported by different actors. For example, an avionics application can use two macro-codes: a macro-code to adapt the behavior of the application to the client's need, and another macro-code to adapt the behavior of the application to the airplane. In this example, the first macro-code may be provided by the Module Supplier, and the second one by the System Integrator.

## Use Case Example

In this example, an application named "Rack Manager" is used to manage an avionics rack (i.e. a set of avionics calculators integrated in a common structure). In case of power supply partial failure, the Rack Manager must stop some calculators installed in the rack. A decision rule is used to choose which calculators have to be stopped.

The Function Supplier provides the Rack Manager application, and he wants to offer to the System Integrator the possibility to define his own decision rule. So the System Integrator writes an adaptation source code to implement his decision rule, and he uses the macro-compiler to generate a compiled file.

The following figure gives an example of such an adaptation source code, using a C-like syntax.

```
// Decision rule to stop calculators
// v1.0

//--- External Variables
bool ground;
struct
{
	bool powered;
	int8 criticity;
}
calculator[1..10];

//--- Local Variables
local int8 i;

//--- Decision Logic
if (ground)
{
	calculator[1].powered = false;
}
else
{
	for (i=1; i<=10; i++)
		if (calculator[i].criticity < 5)
			calculator[i].powered = false;
}
```

**Figure 6. Adaptation Source Code Example**

In this example, the Rack Manager sends two inputs to the adaptation code: the ground condition and the criticity of each calculator. The adaptation code then returns to the Rack Manager the list of calculators to stop (through the values set in “calculator[].powered”).

The compiled file and the Rack Manager application are dataloaded on the target platform. At runtime, the macro-code is used by the Rack Manager to react to a power fall event, as shown in the following figure.

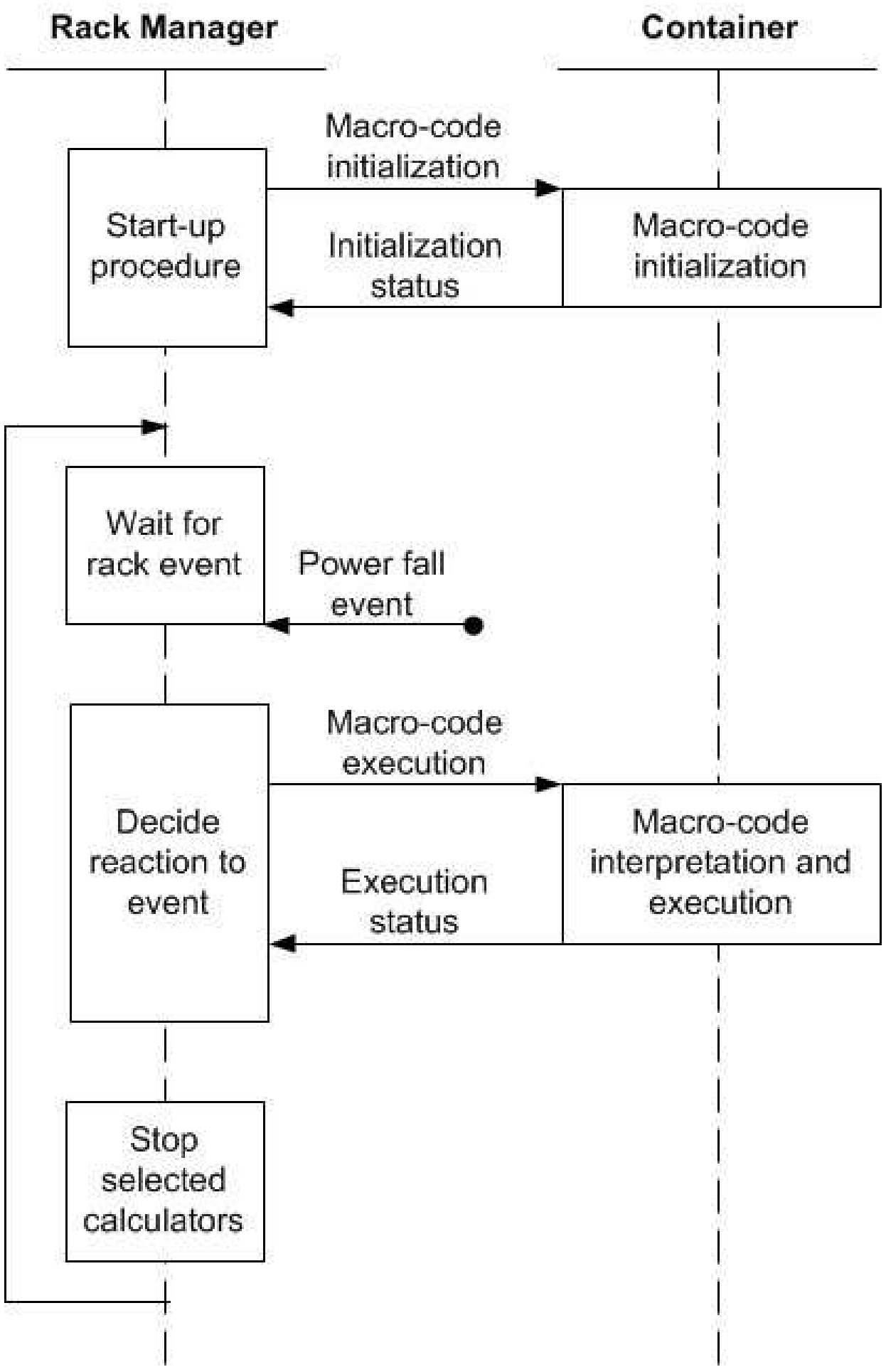

**Figure 7. Use Case Sequence Diagram Example**

## Conclusion

The incremental certification has reduced the granularity of the certification from the module to the partition. In this paper, we have proposed a framework to reduce the certification granularity to a sub-part of software.

With this framework, an avionics application can be adapted to a client or airplane specificities, without re-certifying the entire application. This reduces the cost and the reactivity of software evolutions during the system integration phase.

## Email Addresses

Martin Rayrole:

martin.rayrole@fr.thalesgroup.com

David Faura: david.faura@fr.thalesgroup.com

Marc Gatti: marc-j.gatti@fr.thalesgroup.com